\documentclass[doublecol,linenumbers,amsmath,amssymb]{epl2} 

\usepackage{graphicx}

\title{
Application of Sturm's theorem to  
marginal stable circular orbits of a test body 
in spherically symmetric and static spacetimes  
}
\author{Toshiaki Ono\inst{1}  
\and 
Tomohito Suzuki\inst{1}
\and 
Naomasa Fushimi\inst{1}
\and 
Kei Yamada\inst{2}
\and 
Hideki Asada\inst{1}} 
\institute{
\inst{1} Faculty of Science and Technology, Hirosaki University,
Hirosaki 036-8561, Japan\\
\inst{2} Department of Physics, Kyoto University, 
Kyoto 606-8502, Japan
} 

\abstract{
In terms of Sturm's theorem, 
we reexamine a marginal stable circular orbit (MSCO) such as 
the innermost stable circular orbit (ISCO) of a timelike geodesic 
in any spherically symmetric and static spacetime. 
MSCOs for some of exact solutions to the Einstein's equation 
are discussed. 
Strum's theorem is explicitly applied to the Kottler 
(often called Schwarzschild-de Sitter) spacetime. 
Moreover, we analyze MSCOs 
for a spherically symmetric, static and vacuum solution  
in Weyl conformal gravity. 
}


\begin{document}

\maketitle

\section{Introduction}
In general relativity, the orbital radius 
has a lower bound that is called the innermost stable circular orbit (ISCO). 
The existence of the ISCO 
in the Schwarzschild solution for the Einstein's equation is 
fascinating 
\cite{Text}. 
Moreover, 
ISCOs may play key roles in astrophysics as well as in gravity theory.
For instance, ISCOs are of great importance in gravitational waves 
astronomy \cite{Blanchet}, 
because ISCOs are thought to be the location at the transition from the 
inspiralling phase to the merging one, especially 
when a compact object is orbiting around a massive black hole 
probably located at a galactic center.
Furthermore, in high energy astrophysics, 
ISCOs are related to 
the existence for the inner edge of 
an accretion disk around a black hole 
\cite{Abramowicz}. 
It is thus expected that a measurement of the ISCO radius 
will bring us important information on the strong gravity, 
especially the nonlinear spacetime geometry 
that is beyond the solar-system tests. 
For near-future astrophysical tests of the no-hair theorem 
for black holes, therefore, it is intriguing to study the nature of the ISCOs 
in non-Schwarzschild spacetimes such as black holes 
with electric charges and/or scalar fields 
\cite{no-hair} 
and black holes in modified gravity theories 
\cite{modified}.

It has been thought that unstable circular orbits 
would be generally produced by strong gravity in general relativity. 
This is the case for the Schwarzschild spacetime. 
If the cosmological constant for instance 
is added into the Einstein's equation, 
however, it is not always the case. 
Stuchlik and Hledik \cite{SH} have pointed out that 
the outermost stable circular orbit (OSCO) of a test body is possible 
in the Kottler (often called the Schwarzschild-de Sitter) spacetime 
\cite{Kottler}. 
The ISCO and OSCO are a boundary between a 
stable region and an unstable one. 
Hereafter, we call it a marginal stable circular orbit (MSCO). 
Note that the wording ``marginally stable'' has a different meaning 
in the theory of dynamical systems. 
Hence, this paper prefers to use 
the wording ``marginal stable circular orbits''.
If MSCOs are two and only two in a spacetime, 
one MSCO may correspond to the ISCO and the other 
may correspond to the OSCO. 
Note that the number of MSCOs might be larger than two, 
if such a spacetime geometry is described by 
a very complicated form of the metric. 
For this case, the smallest MSCO may be called the ISCO and 
the largest one may be the OSCO. 

Even counting the number of MSCOs for a non-Schwarzschild metric, 
though it is apparently elementary,  
is not a straightforward task. 
The purpose of this brief paper is to 
reexamine, in terms of Strum's theorem \cite{Algebra},  
MSCOs of a timelike geodesic in any spherically symmetric 
and static spacetime 
that may have a deficit angle. 
MSCOs for some of exact solutions to the Einstein's equation 
are mentioned: 
Schwarzschild, Kottler (often called Schwarzschild-de Sitter) \cite{Kottler}, 
Reissner-Nordstr\"om (RN) \cite{RN}, 
Janis-Newman-Winicour (JNW) spacetimes \cite{JNW}, 
and Barriola-Vilenkin monopole with mass 
that is closely related to a static string \cite{BV}. 
We illustrate how Strum's theorem is explicitly 
applied to the Kottler spacetime. 

Moreover, we  analyze a spherically symmetric, static and 
vacuum solution in Weyl conformal gravity, 
which is an alternative gravity model. 
The exterior metric to a static spherically symmetric 
distribution in Weyl conformal gravity was found 
by Mannheim and Kazanas \cite{MK}. 
This solution expresses a black hole for some parameter range 
and it is a naked singularity for another parameter region. 
Recently, the Weyl-gravity solution has been studied to explain 
rotation curves of galaxy samples without introducing 
a dark matter component \cite{MO}
and the model may provide also a possible size of individual galaxies 
\cite{NB}. 
Moreover, the gravitational lensing by this black hole model 
has been discussed by several groups \cite{Sultana,Cattani}.

Throughout this paper, we use the unit of $G=c=1$.

\section{Application of Strum's theorem 
to exact solutions for the Einstein's equation}

\subsection{Equation for a location of a MSCO}
We consider spherically symmetric and static spacetimes. 
A general form of the line element for these spacetimes is 
\begin{equation}
ds^2 = -A(r) dt^2 + B(r) dr^2 + 
C(r) (d\theta^2 + \sin^2\theta d\phi^2) ,  
\label{ds}
\end{equation}
where we assume $g_{tt} \equiv - A(r) < 0$, 
$g_{rr} \equiv B(r) >0$, 
$g_{\theta\theta} \equiv C(r) >0$, 
namely the spacetime signature as $-, +, +, +$. 
One may often choose $C(r) = r^2$. 
However, it would be more convenient to keep $C(r)$ for some cases 
in the literatures where $C(r) \neq r^2$ \cite{JNW,Chowdhury}, 
because we do not need to add calculations of the coordinate transformation 
for reaching $C(r) = r^2$ for such cases.

A {\it necessary} condition for the existence of a MSCO is known as 
\cite{RZ,BJS}
\begin{equation}
\frac{d}{dr}\left(\frac{1}{A(r)} \right)
\frac{d^2}{dr^2}\left(\frac{1}{C(r)} \right)
- \frac{d}{dr}\left(\frac{1}{C(r)} \right)
\frac{d^2}{dr^2}\left(\frac{1}{A(r)} \right) 
= 0 . 
\label{MSCO}
\end{equation}
This equation determines the radius of the MSCO, 
if the MSCO exists. 
Hereafter, we call Eq. (\ref{MSCO}) MSCO equation. 
Note that the MSCO equation does not contain a metric component $B(r)$ 
nor the constants of motion as $E$ and $L$.

Given a root for Eq. (\ref{MSCO}), 
$E^2$ and $L^2$ are expressed as 
\begin{eqnarray}
E^2 &=& - \frac{1}{\Delta} 
\frac{d}{dr}\left(\frac{1}{C(r)}\right) , 
\label{E^2}
\\
L^2 &=& - \frac{1}{\Delta} 
\frac{d}{dr}\left(\frac{1}{A(r)}\right) , 
\label{L^2}
\end{eqnarray}
where we define a determinant as 
\begin{eqnarray}
\Delta \equiv 
\left|
\begin{array}{cc}
\frac{1}{A(r)} & -\frac{1}{C(r)} \\
\frac{d}{dr}\left(\frac{1}{A(r)}\right) 
& - \frac{d}{dr}\left(\frac{1}{C(r)}\right) 
\end{array}
\right| . 
\label{Delta}
\end{eqnarray}
Any root $r$ for Eq. (\ref{MSCO}) can be substituted into 
Eqs. (\ref{E^2}) and (\ref{L^2}) 
in order to 
see whether 
it satisfies 
the {\it sufficient} condition as $0 \leq E^2 < \infty$ 
and $0 \leq L^2 < \infty$. 
If the sufficient condition is satisfied, 
this $r$ is a MSCO radius, denoted as $r_{MSCO}$. 
If not, the root is unphysical and it must be discarded. 
Note that the expressions of $E^2$ and $L^2$ in Eqs. (\ref{E^2}) 
and (\ref{L^2}) are applicable not only to MSCOs but also 
to any circular orbit.

\subsection{MSCOs in terms of Strum's theorem } 
In this subsection, we apply Strum's theorem to 
some of exact solutions of the Einstein's equation 
in order to classify the number of MSCOs. 

\noindent
{\bf Schwarzschild spacetime:}\\
Let us begin with the Schwarzschild spacetime as 
\begin{equation}
ds^2 = -\left( 1 - \frac{r_g}{r} \right) 
dt^2 
+ \frac{dr^2}{\displaystyle 1 - \frac{r_g}{r}} 
+ r^2 (d\theta^2 + \sin^2\theta d\phi^2) ,  
\label{Sch}
\end{equation}
where the Schwarzschild radius is defined as 
$r_g \equiv 2M$ for the ADM mass $M$. 
For this spacetime metric, Eq. (\ref{MSCO}) becomes linear as 
\begin{equation}
r-3r_g = 0 ,  
\end{equation}
where we assume $r \neq 0$ because $r = 0$ is the spacetime singularity. 
Hence, we obtain $r_{MSCO} = 3 r_g$. 


\noindent{\bf Kottler (Schwarzschild-de Sitter) spacetime:}\\
Next, we consider the Kottler spacetime 
\cite{Kottler}. 
The line element is 
\begin{eqnarray}
ds^2 &=& -\left( 1 - \frac{r_g}{r} - \frac{\Lambda}{3}r^2 \right) 
dt^2 
+ \frac{dr^2}{\displaystyle 1 - \frac{r_g}{r} - \frac{\Lambda}{3}r^2} 
\nonumber\\
&&
+ r^2 (d\theta^2 + \sin^2\theta d\phi^2) , 
\label{ds}
\end{eqnarray}
where $\Lambda$ denotes the cosmological constant. 
Note that this spacetime is not asymptotically flat. 
The timelike geodesics in the Kottler spacetime 
have been often studied by several authors 
\cite{Islam}, 
especially on the periastron shift, 
though none of them has examined the allowed region of the orbits.

For this spacetime, the MSCO equation becomes 
\begin{equation}
\frac83 \Lambda r^4 - 5 r_g \Lambda r^3 - r_g r + 3 r_g^2 = 0 , 
\end{equation}
where we assume $r \neq 0$ and $r \neq \infty$. 
It is convenient to use the normalized variables in terms of 
the Schwarzschild radius $r_g$. 
We define $x \equiv r / r_g$ and 
$\lambda \equiv \Lambda r_g^2 / 3$.  
The above quartic equation becomes 
\begin{equation}
8 \lambda x^4 - 15 \lambda x^3 - x + 3 = 0 . 
\label{x4}
\end{equation}

The remaining task is to investigate positive zeros for 
this quartic equation 
as a necessary condition for MSCOs. 
A general formula for solving a quartic equation such 
as Cardano's one gives a complicated form of the root. 
Hence, it would be preferred to use an alternative method.

{\it Strum's theorem} \cite{Algebra}
expresses the number of distinct real roots of a polynomial $p$ 
located in an interval in terms of the number of changes of signs 
of the values of the Strum's sequence 
(called also the Sturm's chain) 
at the bounds of the interval.

{\it Sturm's theorem}: 
Let $p(r)$ denote a polynomial. 
Applying Euclid's algorithm to $p(r)$ and its derivative, 
Sturm's sequence is constructed as 
\begin{eqnarray}
p_0(r) &\equiv& p(r) , 
\nonumber\\
p_1(r) &\equiv& p_0^{\prime}(r) , 
\nonumber\\
p_2(r) &\equiv& p_1(r) q_0(r) - p_0(r) , 
\nonumber\\
p_3(r) &\equiv& p_2(r) q_1(r) - p_1(r) , 
\nonumber\\
\cdots & & \cdots 
\nonumber\\
0 &=& p_{n}(r) q_{n-1}(r) - p_{n-1}(r) , 
\end{eqnarray}
where $q_i(r)$ is the quotient of $p_{i}(r)$ by $p_{i+1}(r)$. 
Let $\mathcal{V}(a)$ denote the number of the sign changing 
(ignoring zeros) in the Sturm's sequence at $r=a$. 
Then, $\mathcal{V}(a) - \mathcal{V}(b)$ gives the number of 
distinct roots of $p(r)$ between $a$ and $b$, where $a < b$.

Strum's sequence for Eq. (\ref{x4}) in the Kottler spacetime is obtained as 
\begin{eqnarray}
f_0 (r) &=&  8 \lambda r^4 - 15 r_g \lambda r^3 - r_g^3 r + 3 r_g^4 , \\
f_1 (r) &=&  32 \lambda r^3 - 45 r_g \lambda r^2 - r_g^3 , \\
f_2 (r) &=&  \frac{3}{128} r_g^2 ( 225 \lambda r^2 + 32 r_g r - 123 r_g^2) , \\
f_3 (r) &=&  \frac{128 r_g^2}{50625 \lambda} 
\nonumber\\
&& 
\times \left[ - 2 (128 + 4725 \lambda) r + 3 r_g (328 + 3375 \lambda) \right] , 
\nonumber\\
&& 
\\
f_4 (r) &=& - \frac{151875 r_g^4 \lambda}{512 (128 + 4725 \lambda)^2} 
(4 - 27 \lambda) (16 - 16875 \lambda) ,
\nonumber\\
&& 
\end{eqnarray}
where $f_4 (r)$ is a constant for the present case. 
Henceforth, 
we focus on non-Schwarzschild case ($\lambda \neq 0$).

First, we consider the case of $\lambda > 0$, for which we have 
\begin{eqnarray}
&&\mbox{sgn} [f_0 (0)] > 0 , 
\nonumber\\
&&\mbox{sgn} [f_1 (0)] < 0 , 
\nonumber\\
&&\mbox{sgn} [f_2 (0)] < 0 , 
\nonumber\\
&&\mbox{sgn} [f_3 (0)] > 0 , 
\nonumber\\
&&\mbox{sgn} [f_4 (0)] 
= \mbox{sgn} [-\lambda(4 - 27 \lambda) (16 - 16875 \lambda)] , 
\end{eqnarray}
and 
\begin{eqnarray}
&&\mbox{sgn} [f_0 (\infty)] > 0 , 
\nonumber\\
&&\mbox{sgn} [f_1 (\infty)] > 0 , 
\nonumber\\
&&\mbox{sgn} [f_2 (\infty)] > 0 , 
\nonumber\\
&&\mbox{sgn} [f_3 (\infty)] < 0 , 
\nonumber\\
&&\mbox{sgn} [f_4 (\infty)] 
= \mbox{sgn} [-\lambda(4 - 27 \lambda) (16 - 16875 \lambda)] , 
\end{eqnarray}
where $\mbox{sgn}$ denotes the sign function.

(1) If ${16}/{16875} < \lambda < {4}/{27}$, 
the number of the sign changing is $\mathcal{V}(0) = 2$ and 
$\mathcal{V}(\infty) = 2$, so that 
there cannot exist any positive root. 
(2) If $0<\lambda<{16}/{16875}$ or $\lambda>{4}/{27}$, 
the sign changing number is 
$\mathcal{V} (0) = 3$ and 
$\mathcal{V} (\infty) = 1$, which lead to two positive roots. 
(2a) If $\lambda \geq {4}/{27}$, however, $L^2 \leq 0$, 
which means no MSCO. 
(2b) On the other hand, both $E^2 >0$ and $L^2 >0$ are satisfied 
by $0<\lambda<{16}/{16875}$, which means two MSCOs. 

Next, we consider $\lambda < 0$. 
Then, we have 
\begin{eqnarray}
&&\mbox{sgn} [f_0 (0)] > 0 , 
\nonumber\\
&&\mbox{sgn} [f_1 (0)] < 0 , 
\nonumber\\
&&\mbox{sgn} [f_2 (0)] < 0 , 
\nonumber\\
&&\mbox{sgn} [f_3 (0)] =\mbox{sgn}[-(328+3375\lambda)] , 
\nonumber\\
&&\mbox{sgn} [f_4 (0)] > 0 , 
\end{eqnarray}
and 
\begin{eqnarray}
&&\mbox{sgn} [f_0 (\infty)] < 0 , 
\nonumber\\
&&\mbox{sgn} [f_1 (\infty)] < 0 , 
\nonumber\\
&&\mbox{sgn} [f_2 (\infty)] < 0 , 
\nonumber\\
&&\mbox{sgn} [f_3 (\infty)] =\mbox{sgn}[128+4725\lambda] , 
\nonumber\\
&&\mbox{sgn} [f_4 (\infty)] > 0 . 
\end{eqnarray}

Regardless of $\mbox{sgn} [f_3 (0)]$ and $\mbox{sgn} [f_3 (\infty)]$, 
we find $\mathcal{V} (0) = 2$ and 
$\mathcal{V} (\infty) = 1$, which lead to a single positive root. 
One can also see that it satisfies $E^2 > 0$ and $L^2 > 0$. 
If $\lambda < 0$, then, 
there always exists a single MSCO corresponding to the ISCO.

Strum's theorem (as the necessary condition) and 
the positive $E^2$ and $L^2$ (as the sufficient condition) tell 
that there are four cases: 

\noindent 
Case 1: $\lambda =0$. There is the single positive zero 
corresponding to the ISCO. This is in agreement with 
the Schwarzschild case. 

\noindent 
Case 2: $0< \lambda < 16/16875$. 
We have two positive zeros, namely two MSCOs,  
where one is corresponding to the ISCO and the other is the OSCO. 

\noindent 
Case 3: $16/16875 \leq \lambda$. 
There is no MSCO 
(after the ISCO and the OSCO merge at $\lambda = 16/16875$). 
This implies that every circular orbit becomes unstable for this case. 
This number $16/16875$ was found by Stuchlik and Hledik \cite{SH}, 
where this is expressed as $y = 12/15^4$ in Eq. (24) of their paper. 

\noindent 
Case 4: $\lambda < 0$ (anti-de Sitter case). 
There is only the single MSCO.  


\noindent{\bf Other examples in general relativity:}\\ 
Let us briefly mention MSCOs in 
Reissner-Nordstr\"om (RN) \cite{RN}, 
Janis-Newman-Winicour (JNW) spacetimes \cite{JNW}, 
and Barriola-Vilenkin monopole with mass \cite{BV}. 
MSCO equations are cubic, quadratic and linear, respectively, 
though at first glance the JNW case appears to lead to 
a higher order polynomial \cite{Chowdhury}.  
Therefore, Strum's theorem is not needed for these cases.

\noindent{\bf Spherically symmetric, static and vacuum solution 
in Weyl conformal gravity:}\\ 
Finally, we study a spherically symmetric, static and vacuum solution 
in Weyl conformal gravity \cite{MK} as an alternative 
gravity model. 
The exterior spacetime metric in vacuum regions 
was found by Mannheim and Kazanas 
\cite{MK} as 
\begin{eqnarray}
ds^2&=&-A(r) dt^2+\frac{1}{A(r)} dr^2 
+ r^2(d\theta^2+\sin^2\theta d\phi^2), \nonumber\\
A(r)&=&1 - 3 \beta\gamma - \frac{\beta (2-3\beta\gamma)}{r} 
+ \gamma r -k r^2 ,
\label{le-0}
\end{eqnarray}
where $\beta$, $\gamma$ and $k$ are the integration constants 
to the vacuum equation in Weyl conformal gravity \cite{MK}. 
The $k$ parameter may play the same role as the cosmological 
constant in the Kottler spacetime. 

It is more convenient to rearrange this original form of the metric 
into another one \cite{Cattani} as 
\begin{eqnarray}
ds^2&=&-A(r) dt^2+\frac{1}{A(r)} dr^2 
+ r^2(d\theta^2+\sin^2\theta d\phi^2), \nonumber\\
A(r)&=&\alpha-\frac{2m}{r}+\gamma r -k r^2 ,
\label{le}
\end{eqnarray}
where we defined 
$\alpha \equiv \sqrt{1 - 6 m \gamma}$, and 
$m \equiv \beta (2 - 3 \beta\gamma)/2$. 
The negative mass case prohibits a circular orbit because of 
its repulsive force. 
This paper thus focuses on $m > 0$ to study MSCOs. 

In this paper, both cases of $\gamma \geq 0$ and $\gamma < 0$ 
are considered. 
When we wish to discuss a possible explanation 
of galactic rotation curves in Weyl gravity, 
$\gamma > 0$ is preferred \cite{MK,MO}. 
On the other hand, $\gamma$ parameter cannot be 
constrained by galactic rotation curves, 
if one takes the standard cosmological point of view 
that galactic rotation curves can be explained 
by introducing a dark matter component. 
In this stance, therefore, 
the sign of $\gamma$ parameter can be uncertain at present. 

Let us begin with the most general case ($k \neq 0$ and $\gamma \neq 0$). 
Straightforward calculations show that 
MSCO equation becomes quintic in $r$, where the fifth-order term 
is due to a coupling between $k$ and $\gamma$ parameters. 
The quintic equation is not algebraically tractable. 
Moreover, roughly speaking, $r^2$ terms that may play a role 
at cosmological distances 
are negligible at smaller scales such as galactic and stellar sizes 
compared with the linear terms. 
In this paper, therefore, we focus on $k = 0$ and $\gamma \neq 0$ 
in order to neglect the $r^2$ terms in the metric. 
This case in Weyl conformal gravity may be used to discuss 
a modification to a stellar-mass black hole model, 
because $k$ can be effective only at cosmological distances 
as mentioned above. 
See also the metric for Kottler solution. 

Interestingly, Eq. (\ref{le}) with $k = 0$ and $\gamma \neq 0$ 
accidentally coincides with an exact solution of Einstein equation 
with the quintessential matter surrounding a black hole 
(often called Kiselev black hole \cite{Kiselev}), 
if the parameter in Kiselev black hole can be adjusted suitably 
($3 w_n+1 = -1$). 
Note that the parameters of Weyl-gravity black holes 
and Kiselev ones have very different physical origins: 
The parameters in Weyl-gravity black holes are the integration constants 
coming from a higher derivative theory, while 
those in Kiselev black holes are due to 
the equation of state for a dark energy component, 
more precisely the potential form that the quintessence field obeys.


According to Eqs. (\ref{E^2}) and (\ref{L^2}), 
$E^2 > 0$ and $L^2 > 0$ for this spacetime metric are 
\begin{equation}
\gamma r^2 + 2 \sqrt{1-6m} r - 6m > 0 ,
\end{equation}
and
\begin{equation}
\frac{\gamma r^2 + 2m}{\gamma r^2 + 2 \sqrt{1-6m} r - 6m} > 0 , 
\end{equation}
respectively.

If $\gamma > 0$, $A(r) = 0$ gives the black horizon radius as 
\begin{equation}
r_{H} = 
\frac{\sqrt{1 + 2 m \gamma} - \sqrt{1 - 6 m \gamma}}{2 \gamma} . 
\end{equation}
If $\gamma < 0$, $A(r) = 0$ gives the black horizon radius as 
\begin{eqnarray}
r_{H in} &=& 
- \frac{\sqrt{1 - 6 m \gamma} + \sqrt{1 + 2 m \gamma}}{2 \gamma} , 
\\
r_{H out} &=& 
- \frac{\sqrt{1 - 6 m \gamma} - \sqrt{1 + 2 m \gamma}}{2 \gamma} , 
\end{eqnarray}
where the former and the latter correspond to 
the inner horizon and the outer one, respectively. 
Therefore, the present metric is a black hole solution if and only if 
\begin{equation}
- \frac{1}{2 m} < \gamma < \frac{1}{6 m} . 
\label{BH}
\end{equation}

The MSCO equation for $k = 0$ and $\gamma \neq 0$ becomes 
\begin{eqnarray}
&&
\gamma^2 r^4 + 3 \gamma \sqrt{1 - 6 m \gamma} r^3 - 12 m \gamma r^2 
\nonumber\\
&& 
+ 2 m \sqrt{1 - 6 m \gamma} r - 12 m^2 = 0 .
\end{eqnarray}
Let us define 
\begin{eqnarray}
\bar{\gamma} &\equiv& 2 m \gamma, 
\\
\bar{r} &\equiv& \frac{r}{2 m} , 
\end{eqnarray}
so that the Sturm's sequences can be in a simpler form as 
\begin{eqnarray}
f_0 (r) &=& 
- \bar{\gamma}^2 \bar{r}^4 
- 3 \bar{\gamma} \sqrt{1 - 3 \bar{\gamma}} \bar{r}^3 
+ 6 \bar{\gamma} \bar{r}^2 
\nonumber\\
&&
- \sqrt{1 - 3 \bar{\gamma}} \bar{r} + 3 , \\
f_1 (r) &= &
- 4 \bar{\gamma}^2 \bar{r}^3 
- 9 \sqrt{1 - 3 \bar{\gamma}} \bar{\gamma} \bar{r}^2 
+ 12 \bar{\gamma} \bar{r} 
\nonumber\\
&&
- \sqrt{1 - 3 \bar{\gamma}} , \\
f_2 (r) &=& 
- \frac{3}{16} ( 9 - 11 \bar{\gamma} ) \bar{r}^2 
+ 3 \sqrt{1 - 3 \bar{\gamma}} \bar{r} 
\nonumber\\
&&
- \frac{3}{16 \bar{\gamma}} ( 1 + 13 \bar{\gamma} ) , \\
f_3 (r) &=& 
\frac{32 \bar{\gamma}}{( 9 - 11 \bar{\gamma} )^2} 
( 9 - 78 \bar{\gamma} + 25 \bar{\gamma}^2 ) \bar{r} 
\nonumber\\
&&
- \frac{64 \sqrt{1 - 3 \bar{\gamma}} \bar{\gamma}}{( 9 - 11 \bar{\gamma} )^2} 
( 19 - 9 \bar{\gamma} ) , \\
f_4 (r) &=& 
\frac{3 ( 9 - 11 \bar{\gamma} )^2}
{16 \bar{\gamma} ( 3 - 25 \bar{\gamma} )^2 ( - 3 + \bar{\gamma} )^2}
\nonumber\\
&&
\times 
( 1 + \bar{\gamma} ) ( 1 + 90 \bar{\gamma} - 23 \bar{\gamma}^2 ) ,
\end{eqnarray}
where $f_4(r)$ is a constant. 
Eq. (\ref{BH}) is rewritten as 
\begin{equation}
- 1 < \bar{\gamma} < \frac13 . 
\end{equation}

\noindent 
Case 1: $\bar{\gamma} \ge 0$. 
The sign of the Sturm's sequence for this case is 
\begin{eqnarray}
\mbox{sgn} [f_0 (0)] &>& 0 , \\
\mbox{sgn} [f_1 (0)] &<& 0 , \\
\mbox{sgn} [f_2 (0)] &<& 0 , \\
\mbox{sgn} [f_3 (0)] &<& 0 , \\
\mbox{sgn} [f_4 (0)] 
&=& \mbox{sgn} [1 + 90 \bar{\gamma} - 23 \bar{\gamma}^2] , 
\end{eqnarray}
and 
\begin{eqnarray}
\mbox{sgn} [ f_0 (\infty) ] &<& 0 , \\ 
\mbox{sgn} [ f_1 (\infty) ] &<& 0 , \\ 
\mbox{sgn} [ f_2 (\infty) ] 
&=& 
\mbox{sgn} [ - ( 9 - 11 \bar{\gamma} ) ] , \\ 
\mbox{sgn} [ f_3 (\infty) ] &=& 
\mbox{sgn} [ 9 - 78 \bar{\gamma} + 25 \bar{\gamma}^2 ], \\ 
\mbox{sgn} [ f_4 (\infty) ] &=& 
\mbox{sgn} [1 + 90 \bar{\gamma} - 23 \bar{\gamma}^2] .
\label{weyl1}
\end{eqnarray}

There is a constraint as $\bar{\gamma} < 1/3$, 
because there exists the $\sqrt{1 - 3\bar{\gamma}}$ term in the metric. 
Hence, we can show 
\begin{eqnarray}
\mbox{sgn} [ f_2 (\infty) ] &<& 0 , 
\\
\mbox{sgn} [ f_4 (0) ] &=& \mbox{sgn} [ f_4 (\infty) ] 
\nonumber\\
&>& 0 . 
\end{eqnarray}

Regardless of $\mbox{sgn} [ f_3 (\infty) ] $, 
hence, $\mathcal{V} (0)=2$ and $\mathcal{V} (\infty)=1$. 
Therefore, there exists only the single positive root. 
Moreover, it satisfies $E^2 > 0$ and $L^2 > 0$. 
Hence, there is the only one MSCO, namely ISCO. 

\noindent
Case 2: $\bar{\gamma} < 0$. 
This case is rather tricky. Non-trivial calculations are 
needed as shown below. 
The sign of the Sturm's sequence for this case is 
\begin{eqnarray}
\mbox{sgn} [ f_0 (0) ] &>& 0 , \\ 
\mbox{sgn} [ f_1 (0) ] &<& 0 , \\ 
\mbox{sgn} [ f_2 (0) ] &=& \mbox{sgn} [ 1 + 13 \bar{\gamma} ] , \\ 
\mbox{sgn} [ f_3 (0) ] &>& 0 , \\ 
\mbox{sgn} [ f_4 (0) ] &=& 
\mbox{sgn} 
[ - ( 1 + \bar{\gamma} ) ( 1 + 90 \bar{\gamma} - 23 \bar{\gamma}^2 ) ] , 
\end{eqnarray}
and 
\begin{eqnarray}
\mbox{sgn} [ f_0 (\infty) ] &<& 0 , \\ 
\mbox{sgn} [ f_1 (\infty) ] &<& 0 , \\ 
\mbox{sgn} [ f_2 (\infty) ] &<& 0 , \\ 
\mbox{sgn} [ f_3 (\infty) ] &=& 
\mbox{sgn} [ - ( 9 - 78 \bar{\gamma} + 25 \bar{\gamma}^2 ) ] , \\ 
\mbox{sgn} [ f_4 (\infty) ] &=& 
\mbox{sgn} 
[ - ( 1 + \bar{\gamma} ) ( 1 + 90 \bar{\gamma} - 23 \bar{\gamma}^2 ) ] .
\end{eqnarray}

For $- 1 < \bar{\gamma} < 0$, one can show that 
\begin{eqnarray}
{\rm sgn} [ - ( 1 + \bar{\gamma} ) ] &<& 0 , \\
{\rm sgn} [ f_3 (\infty) ] &<& 0 .
\end{eqnarray}
Regardless of the sign of $ f_2 (0)$, therefore, 
the number of the positive roots changes at 
\begin{eqnarray}
\bar{\gamma} = \frac{45 - 32 \sqrt{2}}{23} .
\end{eqnarray}

As a result, there are three types for $- 1 < \bar{\gamma} < 0$:\\
\noindent 
(2-1) 
If $(45 - 32 \sqrt{2})/23 < \bar{\gamma} < 0$, 
there are three positive zeros.
However, one of them is located outside the outer horizon, 
while the others are placed between the two horizons 
and they satisfy both $E^2 > 0$ and $L^2 > 0$. 
Namely, there are two MSCOs.\\ 
\noindent 
(2-2) 
If $\bar{\gamma} = (45 - 32 \sqrt{2})/23$, 
the ISCO and the OSCO merge.\\
\noindent 
(2-3)
If $- 1 < \bar{\gamma} < (45 - 32 \sqrt{2})/23$, 
the MSCO equation has one positive root.
However, it does not satisfy $E^2 > 0$. 

\begin{figure}
\begin{center}
\includegraphics[width=8cm]{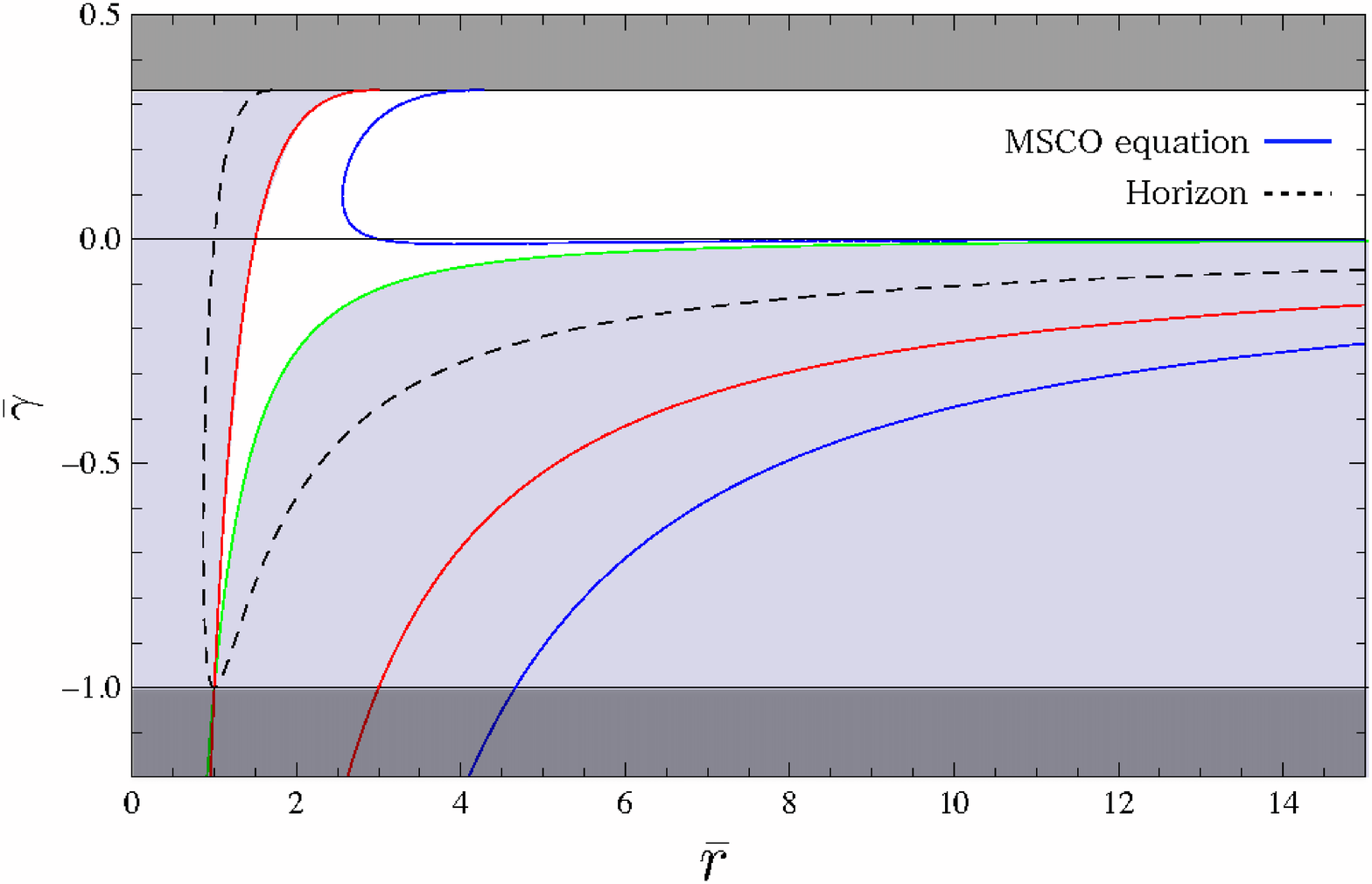}
\includegraphics[width=8cm]{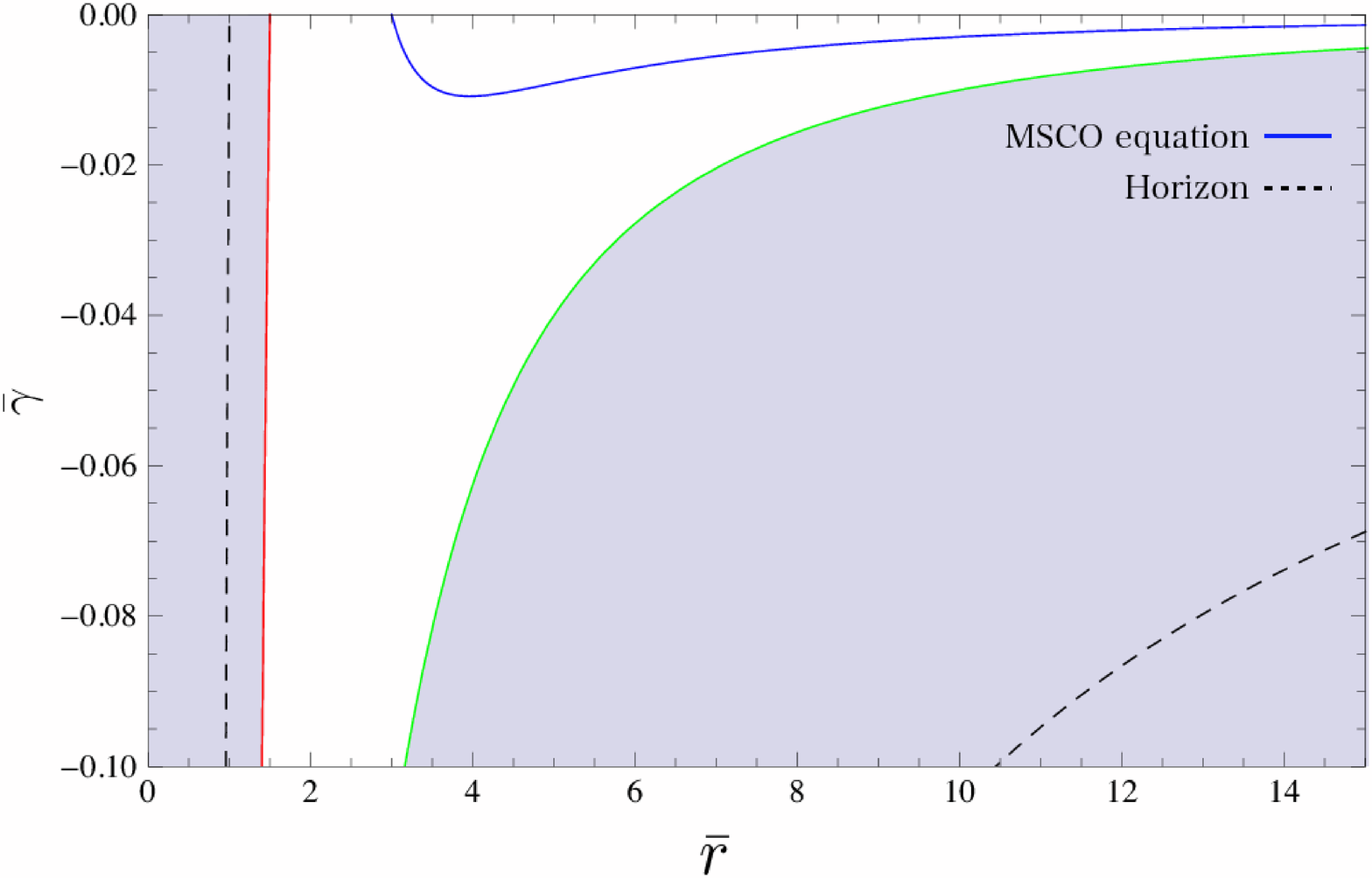}
\caption{
$\bar{\gamma}$-dependence of the MSCO radius. 
The horizontal axis denotes the MSCO radius $r_{MSCO}$ 
as a root for the MSCO equation. 
The vertical axis denotes $\bar{\gamma}$ parameter. 
The shaded parts denote prohibited regions where 
$E^2 < 0$, $L^2 < 0$, $\bar{\gamma} < -1$ or $\bar{\gamma} > 1/3$. 
The dashed curve in the shaded regions denotes the horizon location 
for a given $\bar{\gamma}$ value. 
Top: $\bar{\gamma} \in [-1.2, 0.5]$. 
Bottom: $\bar{\gamma} \in [-0.1, 0]$ 
for a zoom-up figure of the Case 2.
}
\label{f1}
\end{center}
\end{figure}

Bringing the Case 1 (corresponding to $\gamma > 0$) 
and the Case 2 (corresponding to $\gamma < 0$) together, 
Figure \ref{f1} shows the dependence of the MSCO radius on $\bar{\gamma}$.

\section{Conclusion}
We reexamined, in terms of Strum's theorem, 
MSCOs of a timelike geodesic in any spherically symmetric 
and static spacetime 
that may have a deficit angle. 
MSCOs for some of exact solutions to the Einstein's equation 
were mentioned: 
Schwarzschild, Kottler (often called Schwarzschild-de Sitter) \cite{Kottler}, 
Reissner-Nordstr\"om (RN) \cite{RN}, 
Janis-Newman-Winicour (JNW) spacetimes \cite{JNW}, 
and Barriola-Vilenkin monopole with mass \cite{BV}. 
We illustrated how Strum's theorem is explicitly applied 
to the Kottler case. 
This suggests that Strum's theorem is widely applicable for 
classifying MSCOs for some spacetime, 
when the MSCO equation under study is equivalent to a polynomial. 

Moreover, we analyzed MSCOs for a spherically symmetric, static and vacuum 
black hole solution  
in Weyl conformal gravity, where we neglected the square terms 
by assuming $k = 0$. 
MSCO equation becomes quintic for a more general black hole solution 
in Weyl gravity with non-zero $k$ parameter. 
Numerical investigations are needed to classify MSCOs 
in this quintic case. 
It is left as a future work.

\acknowledgments
We wish to thank 
L. Rezzolla, T. Futamase, S. Yoshida, T. Kitamura, and Y. Sendouda 
for useful discussions. 
This work was supported in part 
by JSPS Grant-in-Aid for JSPS Fellows, No. 24108 (K.Y.), 
and JSPS Grant-in-Aid for Scientific Research, 
No. 26400262 (H.A.) and No. 15H00772 (H.A.)

\end{document}